\pdfoutput=1
\documentclass[12pt,a4paper]{article}

\usepackage{ifthen} 
\newboolean{pdflatex}
\setboolean{pdflatex}{true} 

\newboolean{articletitles}
\setboolean{articletitles}{true} 

\newboolean{uprightparticles}
\setboolean{uprightparticles}{false} 

\def\paperauthors{LHCb collaboration} 
\def\paperasciititle{First evidence of CP violation in beauty baryon to charmonium decays } 
\def\papertitle{First evidence of ${C\!P}$ violation in beauty baryon to charmonium decays} 
\def\paperkeywords{{High Energy Physics}, {LHCb}, {CP violation}} 
\def\papercopyright{\the\year\ CERN for the benefit of the LHCb collaboration} 
\def\paperlicence{CC BY 4.0 licence}
\def\paperlicenceurl{https://creativecommons.org/licenses/by/4.0/}

\newif\ifEnableSectionTOCLinks
\EnableSectionTOCLinksfalse 

\usepackage[top=1in, bottom=1.25in, left=1in, right=1in]{geometry}

\usepackage{microtype}
\usepackage{lineno}  
\usepackage{xspace} 
\usepackage{caption} 

\usepackage{graphicx}  
\usepackage{color}
\usepackage{colortbl}
\graphicspath{{./figs/}} 

\usepackage{amsmath} 
\usepackage{amssymb}
\usepackage{amsfonts}
\usepackage{upgreek} 

\newcommand*\patchAmsMathEnvironmentForLineno[1]{%
\expandafter\let\csname old#1\expandafter\endcsname\csname #1\endcsname
\expandafter\let\csname oldend#1\expandafter\endcsname\csname
end#1\endcsname
 \renewenvironment{#1}%
   {\linenomath\csname old#1\endcsname}%
   {\csname oldend#1\endcsname\endlinenomath}%
}
\newcommand*\patchBothAmsMathEnvironmentsForLineno[1]{%
  \patchAmsMathEnvironmentForLineno{#1}%
  \patchAmsMathEnvironmentForLineno{#1*}%
}
\AtBeginDocument{%
\patchBothAmsMathEnvironmentsForLineno{equation}%
\patchBothAmsMathEnvironmentsForLineno{align}%
\patchBothAmsMathEnvironmentsForLineno{flalign}%
\patchBothAmsMathEnvironmentsForLineno{alignat}%
\patchBothAmsMathEnvironmentsForLineno{gather}%
\patchBothAmsMathEnvironmentsForLineno{multline}%
\patchBothAmsMathEnvironmentsForLineno{eqnarray}%
}

\usepackage{hyperxmp}

\usepackage[pdftex,
            pdfauthor={\paperauthors},
            pdftitle={\paperasciititle},
            pdfkeywords={\paperkeywords},
            pdfcopyright={Copyright (C) \papercopyright},
            pdflicenseurl={\paperlicenceurl}]{hyperref}

\usepackage[colorinlistoftodos,textsize=scriptsize]{todonotes}

\usepackage[bottom,flushmargin,hang,multiple]{footmisc}

\usepackage[all]{hypcap} 

\usepackage{xspace} 
\usepackage{upgreek}

\def\lhcb   {\mbox{LHCb}\xspace}

\def\MagUp {\mbox{\em Mag\kern -0.05em Up}\xspace}

\ifthenelse{\boolean{uprightparticles}}%
{

 \def\Pmu         {\ensuremath{\upmu}\xspace}

 \def\Ppi         {\ensuremath{\uppi}\xspace}

 \def\Ppsi        {\ensuremath{\uppsi}\xspace}

 \def\PDelta      {\ensuremath{\Delta}\xspace}                 
 \def\PXi         {\ensuremath{\Xi}\xspace}                 
 \def\PLambda     {\ensuremath{\Lambda}\xspace}                 
 \def\PSigma      {\ensuremath{\Sigma}\xspace}                 
 \def\POmega      {\ensuremath{\Omega}\xspace}                 
 \def\PUpsilon    {\ensuremath{\Upsilon}\xspace}
 \let\oldPi\Pi
 \def\PPi         {\ensuremath{\oldPi}\xspace}

 \def\PB      {\ensuremath{\mathrm{B}}\xspace}                 
 \def\PD      {\ensuremath{\mathrm{D}}\xspace}                 
 \def\PJ      {\ensuremath{\mathrm{J}}\xspace}                 
 \def\PK      {\ensuremath{\mathrm{K}}\xspace}                 
 \def\Pb      {\ensuremath{\mathrm{b}}\xspace}                 
 \def\Pc      {\ensuremath{\mathrm{c}}\xspace}                 
 \def\Pd      {\ensuremath{\mathrm{d}}\xspace}                 

 \def\Pp      {\ensuremath{\mathrm{p}}\xspace}                 

 \def\Ps      {\ensuremath{\mathrm{s}}\xspace}                 
                  
 \def\Pu      {\ensuremath{\mathrm{u}}\xspace}                 
 \def\thebaroffset{0.0em}
}
{

 \def\Pmu         {\ensuremath{\mu}\xspace}

 \def\Ppi         {\ensuremath{\pi}\xspace}

 \def\Ppsi        {\ensuremath{\psi}\xspace}                 
                  
 \mathchardef\PDelta="7101
 \mathchardef\PXi="7104
 \mathchardef\PLambda="7103
 \mathchardef\PSigma="7106
 \mathchardef\POmega="710A
 \mathchardef\PUpsilon="7107
 \mathchardef\PPi="7105
 \def\PB      {\ensuremath{B}\xspace}                 
 \def\PD      {\ensuremath{D}\xspace}                 
 \def\PJ      {\ensuremath{J}\xspace}                 
 \def\PK      {\ensuremath{K}\xspace}                 
 \def\Pb      {\ensuremath{b}\xspace}                 
 \def\Pc      {\ensuremath{c}\xspace}                 
 \def\Pd      {\ensuremath{d}\xspace}                 

 \def\Pp      {\ensuremath{p}\xspace}                 

 \def\Ps      {\ensuremath{s}\xspace}                 
                  
 \def\Pu      {\ensuremath{u}\xspace}                 
 \def\thebaroffset{0.18em}
}
\newcommand{\offsetoverline}[2][\thebaroffset]{\kern #1\overline{\kern -#1 #2}}%

\makeatletter
\ifcase \@ptsize \relax
  \newcommand{\miniscule}{\@setfontsize\miniscule{4}{5}}
\or
  \newcommand{\miniscule}{\@setfontsize\miniscule{5}{6}}
\or
  \newcommand{\miniscule}{\@setfontsize\miniscule{5}{6}}
\fi
\makeatother

\DeclareRobustCommand{\optbar}[1]{\shortstack{{\miniscule (\rule[.5ex]{1.25em}{.18mm})}
  \\ [-.7ex] $#1$}}

\def\mup        {{\ensuremath{\Pmu^+}}\xspace}
\def\mun        {{\ensuremath{\Pmu^-}}\xspace} 

\def\uquark    {{\ensuremath{\Pu}}\xspace}
\def\uquarkbar {{\ensuremath{\overline \uquark}}\xspace}
\def\uubar     {{\ensuremath{\uquark\uquarkbar}}\xspace}
\def\dquark    {{\ensuremath{\Pd}}\xspace}

\def\squark    {{\ensuremath{\Ps}}\xspace}

\def\cquark    {{\ensuremath{\Pc}}\xspace}
\def\cquarkbar {{\ensuremath{\overline \cquark}}\xspace}
\def\ccbar     {{\ensuremath{\cquark\cquarkbar}}\xspace}
\def\bquark    {{\ensuremath{\Pb}}\xspace}

\def\pion   {{\ensuremath{\Ppi}}\xspace}

\def\pip    {{\ensuremath{\pion^+}}\xspace}
\def\pim    {{\ensuremath{\pion^-}}\xspace}

\def\kaon    {{\ensuremath{\PK}}\xspace}

\def\KorKbar {\kern \thebaroffset\optbar{\kern -\thebaroffset \PK}{}\xspace}

\def\Kp      {{\ensuremath{\kaon^+}}\xspace}
\def\Km      {{\ensuremath{\kaon^-}}\xspace}

\def\KS      {{\ensuremath{\kaon^0_{\mathrm{S}}}}\xspace}

\def\D       {{\ensuremath{\PD}}\xspace}

\def\DorDbar {\kern \thebaroffset\optbar{\kern -\thebaroffset \PD}\xspace}
\def\Dz      {{\ensuremath{\D^0}}\xspace}

\def\Dp      {{\ensuremath{\D^+}}\xspace}
\def\Dm      {{\ensuremath{\D^-}}\xspace}

\def\DpDm    {\ensuremath{\Dp {\kern -0.16em \Dm}}\xspace}

\def\B       {{\ensuremath{\PB}}\xspace}
\def\Bbar    {{\ensuremath{\offsetoverline{\PB}}}\xspace}

\def\BorBbar {\kern \thebaroffset\optbar{\kern -\thebaroffset \PB}\xspace}
\def\Bz      {{\ensuremath{\B^0}}\xspace}
\def\Bzb     {{\ensuremath{\Bbar{}^0}}\xspace}
\def\Bd      {{\ensuremath{\B^0}}\xspace}

\def\BdorBdbar {\kern \thebaroffset\optbar{\kern -\thebaroffset \Bd}\xspace}
\def\Bu      {{\ensuremath{\B^+}}\xspace}

\def\Bp      {{\ensuremath{\Bu}}\xspace}

\def\Bs      {{\ensuremath{\B^0_\squark}}\xspace}

\def\BsorBsbar {\kern \thebaroffset\optbar{\kern -\thebaroffset \Bs}\xspace}

\def\jpsi     {{\ensuremath{{\PJ\mskip -3mu/\mskip -2mu\Ppsi}}}\xspace}

\def\Y#1S{\ensuremath{\PUpsilon{(#1S)}}\xspace}

\def\proton      {{\ensuremath{\Pp}}\xspace}
\def\antiproton  {{\ensuremath{\overline \proton}}\xspace}

\def\Lz          {{\ensuremath{\PLambda}}\xspace}
\def\Lbar        {{\ensuremath{\offsetoverline{\PLambda}}}\xspace}
\def\LorLbar     {\kern \thebaroffset\optbar{\kern -\thebaroffset \PLambda}\xspace}

\def\Xires       {{\ensuremath{\PXi}}\xspace}

\def\Lb           {{\ensuremath{\Lz^0_\bquark}}\xspace}
\def\Lbbar        {{\ensuremath{\Lbar{}^0_\bquark}}\xspace}

\def\Xibz         {{\ensuremath{\Xires^0_\bquark}}\xspace}
\def\Xibm         {{\ensuremath{\Xires^-_\bquark}}\xspace}

\newcommand{\decay}[2]{\mbox{\ensuremath{#1\!\to #2}}\xspace} 

\def\to                 {\ensuremath{\rightarrow}\xspace}

\def\CP                {{\ensuremath{C\!P}}\xspace}

\newcommand{\ACP}{{\ensuremath{{\mathcal{A}}^{\CP}}}\xspace}

\def\AT#1     {\ensuremath{A_{\mathrm{T}}^{#1}}\xspace}           

\def\C#1      {\ensuremath{\mathcal{C}_{#1}}\xspace}                       
\def\Cp#1     {\ensuremath{\mathcal{C}_{#1}^{'}}\xspace}                    
\def\Ceff#1   {\ensuremath{\mathcal{C}_{#1}^{\mathrm{(eff)}}}\xspace}        
\def\Cpeff#1  {\ensuremath{\mathcal{C}_{#1}^{'\mathrm{(eff)}}}\xspace}       
\def\Ope#1    {\ensuremath{\mathcal{O}_{#1}}\xspace}                       
\def\Opep#1   {\ensuremath{\mathcal{O}_{#1}^{'}}\xspace}                    

\newcommand{\nospaceunit}[1]{\ensuremath{\text{#1}}}       
\newcommand{\aunit}[1]{\ensuremath{\text{\,#1}}}       

\newcommand{\tev}{\aunit{Te\kern -0.1em V}\xspace}
\newcommand{\gev}{\aunit{Ge\kern -0.1em V}\xspace}
\newcommand{\mev}{\aunit{Me\kern -0.1em V}\xspace}
\newcommand{\kev}{\aunit{ke\kern -0.1em V}\xspace}
\newcommand{\ev}{\aunit{e\kern -0.1em V}\xspace}
 
\newcommand{\mevc}{\ensuremath{\aunit{Me\kern -0.1em V\!/}c}\xspace}
\newcommand{\gevc}{\ensuremath{\aunit{Ge\kern -0.1em V\!/}c}\xspace}
\newcommand{\mevcc}{\ensuremath{\aunit{Me\kern -0.1em V\!/}c^2}\xspace}
\newcommand{\gevcc}{\ensuremath{\aunit{Ge\kern -0.1em V\!/}c^2}\xspace}

\def\mum  {\ensuremath{\,\upmu\nospaceunit{m}}\xspace}

\def\fb   {\ensuremath{\aunit{fb}}\xspace}
\def\invfb   {\ensuremath{\fb^{-1}}\xspace}

\newcommand{\chisq}{\ensuremath{\chi^2}\xspace}

\def\gsim{{~\raise.15em\hbox{$>$}\kern-.85em
          \lower.35em\hbox{$\sim$}~}\xspace}
\def\lsim{{~\raise.15em\hbox{$<$}\kern-.85em
          \lower.35em\hbox{$\sim$}~}\xspace}

\def\sPlot{\mbox{\em sPlot}\xspace}

\def\pt         {\ensuremath{p_{\mathrm{T}}}\xspace}

\def\evtgen     {\mbox{\textsc{EvtGen}}\xspace}

\def\geant      {\mbox{\textsc{Geant4}}\xspace}

\def\photos     {\mbox{\textsc{Photos}}\xspace}

\def\pythia     {\mbox{\textsc{Pythia}}\xspace}

\def\tell1  {TELL1\xspace}
\def\ukl1   {UKL1\xspace}

\newcommand{\ie}{\mbox{\itshape i.e.}\xspace}

\newcommand{\lhcborcid}[1]{\href{https://orcid.org/#1}{\hspace*{0.1em}\raisebox{-0.45ex}{\includegraphics[width=1em]{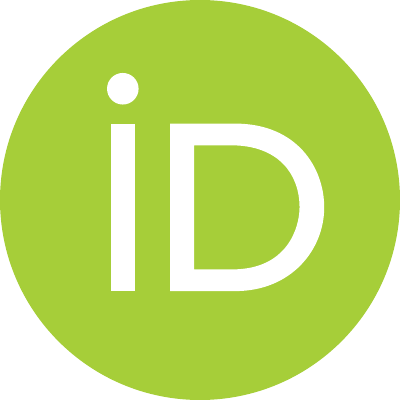}}}}

\hypersetup{
  colorlinks   = true, 
  urlcolor     = blue, 
  linkcolor    = blue, 
  citecolor    = red   
}

\ifEnableSectionTOCLinks
    \usepackage[explicit]{titlesec} 
    
    \let\oldcontentsline\contentsline
    \renewcommand

    \titleformat{\section}{\normalfont\Large\bf}{\hyperlink{tocsection.\thesection}{{\thesection} \parbox[t]{\dimexpr\textwidth-1pc}{#1}}}{1pc}{}

    \titleformat{\subsection}{\normalfont\bf}{\hyperlink{tocsubsection.\thesubsection}{{\thesubsection} \parbox[t]{\dimexpr\textwidth-1pc}{#1}}}{1pc}{}

    \titleformat{name=\section,numberless}[display]{}{}{0pt}{\normalfont\Huge\bfseries #1}
\fi

\usepackage{cite} 
\usepackage{mciteplus}

\usepackage{subcaption}
\usepackage{bm}
\usepackage{multirow}
\usepackage{dcolumn}
\usepackage{placeins}

\newcommand{\varrad}{-0.94}
\newcommand{\varraderr}{0.04}
\newcommand{\varrpiraw}{5.94}
\newcommand{\varrpirawerr}{1.14}
\newcommand{\varrkraw}{0.98\pm0.30}

\newcommand{\varrtoterr}{1.18}
\newcommand{\varrpitpa}{-1.37\pm1.15}

\newcommand{\varrsigsys}{0.016}
\newcommand{\varrbkgsys}{0.012}

\newcommand{\varrtot}{4.03}
\newcommand{\varrresultsys}{0.23}
\newcommand{\varrresulttpa}{-1.37}
\newcommand{\varrresulttpaerr}{1.15}
\newcommand{\varrresulttpasys}{0.21}
\newcommand{\varrcomb}{4.31}
\newcommand{\varrcombstat}{1.06}
\newcommand{\varrcombsys}{0.28}

\def\LbJpsipK      {\ensuremath{\mbox{\decay{\Lb}{\jpsi\proton\Km}}}\xspace}
\def\LbJpsippi      {\ensuremath{\mbox{\Lb\to\jpsi\proton\pim}}\xspace}

\def\ACP         {\ensuremath{{\cal A}_{\CP}}\xspace}
\def\DACP         {\ensuremath{\Delta{\cal A}_{\CP}}\xspace}
\def\Araw         {\ensuremath{{\cal A}_\mathrm{raw}}\xspace}
\def\Atodd        {\ensuremath{\mathcal{A}_{\widehat{T}-\mathrm{odd}}}\xspace}

\makeatletter
\newcolumntype{C}[2]{%
  @{}>{\centering \DC@{int{#1}}{dec{#2}}{}{}{}}c<{}%
}
\makeatother

\begin{document}

\renewcommand{\thefootnote}{\fnsymbol{footnote}}
\setcounter{footnote}{1}



\begin{titlepage}
\pagenumbering{roman}

\vspace*{-1.5cm}
\centerline{\large EUROPEAN ORGANIZATION FOR NUCLEAR RESEARCH (CERN)}
\vspace*{1.5cm}
\noindent
\begin{tabular*}{\linewidth}{lc@{\extracolsep{\fill}}r@{\extracolsep{0pt}}}
\ifthenelse{\boolean{pdflatex}}
{\vspace*{-1.5cm}\mbox{\!\!\!\includegraphics[width=.14\textwidth]{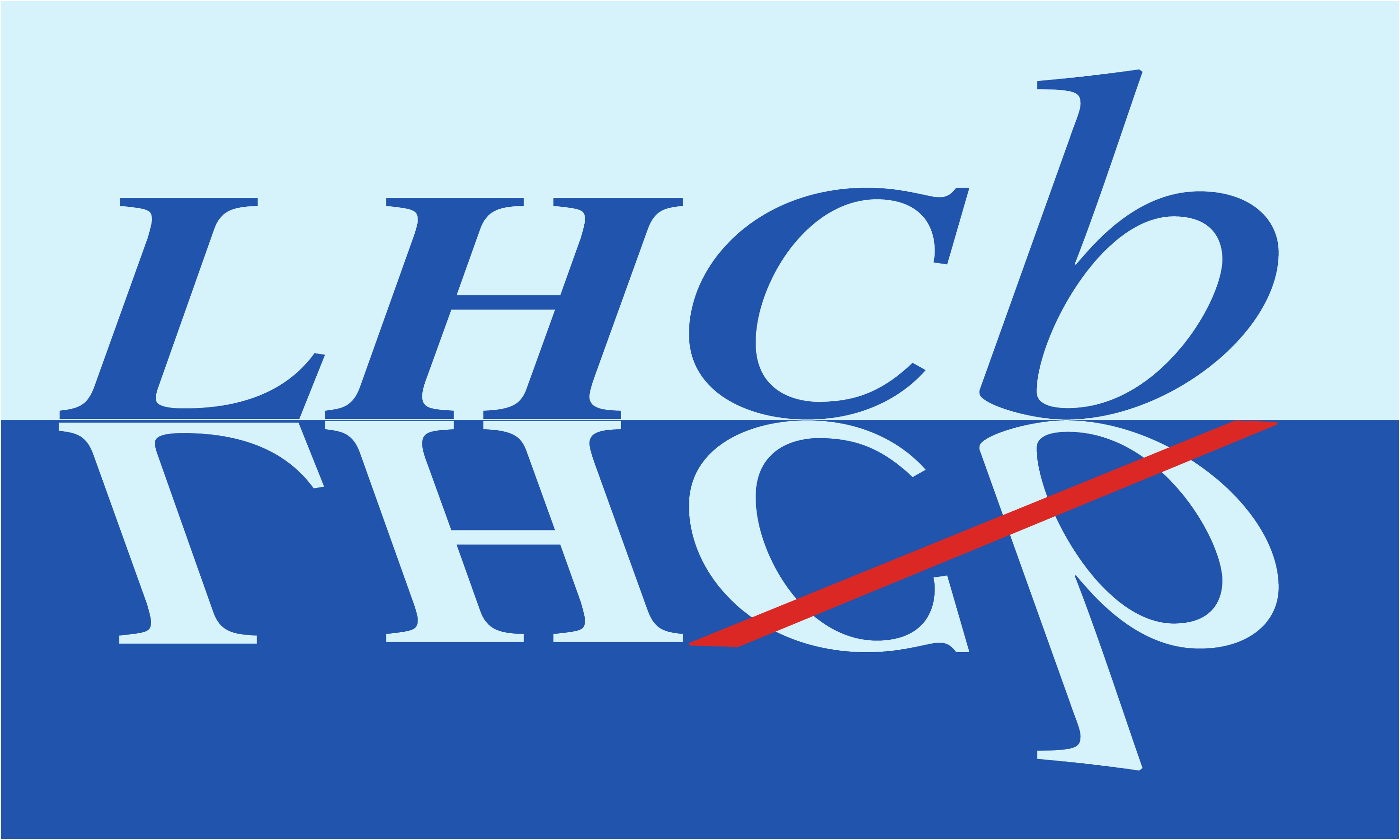}} & &}%
{\vspace*{-1.2cm}\mbox{\!\!\!\includegraphics[width=.12\textwidth]{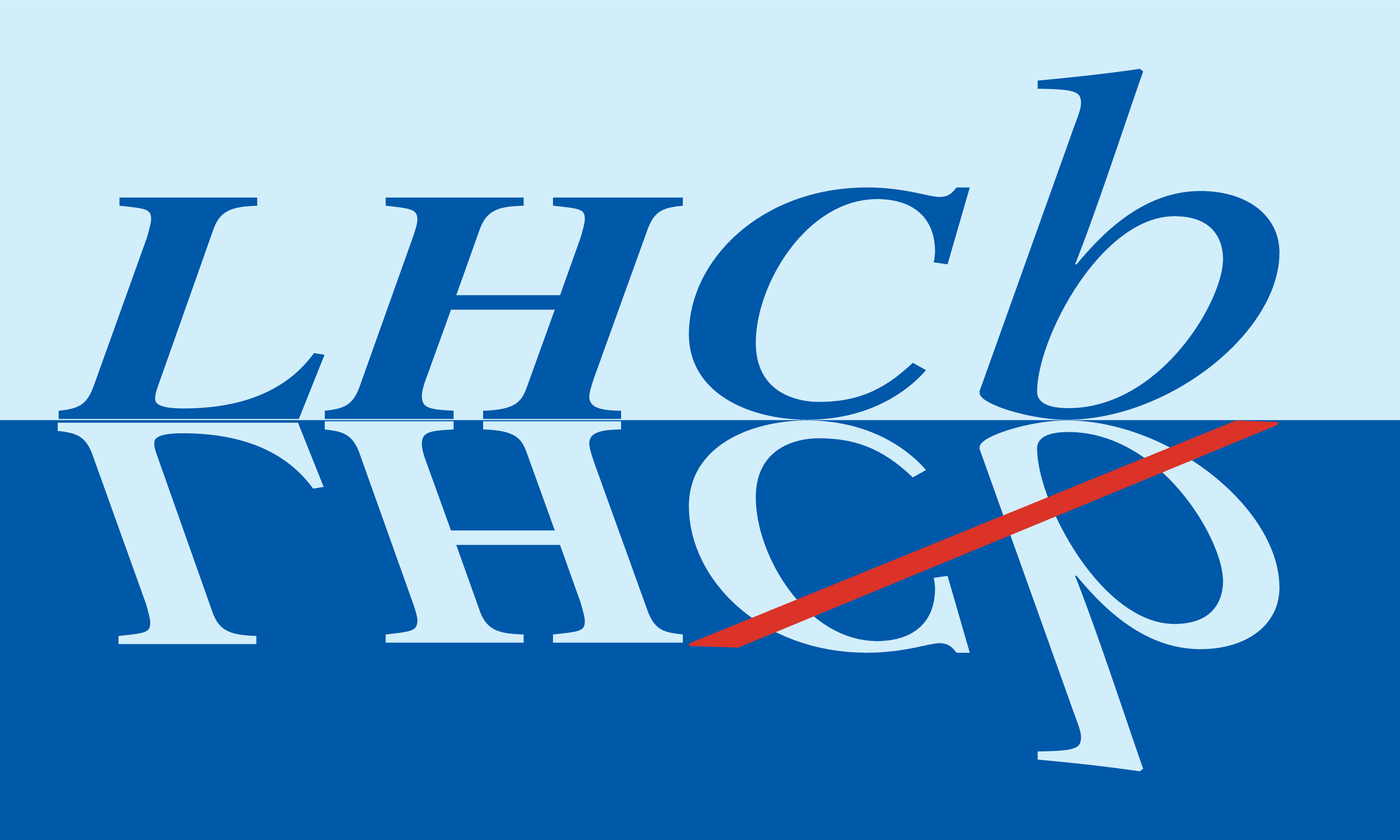}} & &}%
\\
 & & CERN-EP-2025-184 \\  
 & & LHCb-PAPER-2025-021  \\  
 & & 15 February 2026 \\ 
 & & \\
\end{tabular*}

\vspace*{4.0cm}

{\normalfont\bfseries\boldmath\huge
\begin{center}
  \papertitle 
\end{center}
}

\vspace*{2.0cm}

\begin{center}
\paperauthors\footnote{Authors are listed at the end of this paper.}
\end{center}

\vspace{\fill}

\begin{abstract}
  \noindent
  A study of the difference in the $C\!P$ asymmetries between \mbox{\ensuremath{{\mathchar"7103}^0_b \!\rightarrow J\mskip -3mu/\mskip -2mu\psi p \pi^-}}  and  \mbox{\ensuremath{{\mathchar"7103}^0_b \!\rightarrow J\mskip -3mu/\mskip -2mu\psi p K^-}}  decays, $\Delta {\cal A}_{C\!P}$, is performed using proton-proton collision data collected by the LHCb experiment in the years 2015--2018, corresponding to an integrated luminosity of $6\,\mathrm{fb}^{-1}$. This quantity is measured to be \mbox{$ \Delta {\cal A}_{C\!P}=(4.03\pm 1.18\pm 0.23)\%$}, where the first uncertainty is statistical and the second is systematic. When combined with the previous LHCb result, a value of $\Delta {\cal A}_{C\!P} = (4.31 \pm 1.06 \pm 0.28)\%$ is obtained, corresponding to a significance of $3.9\sigma$ against the $C\!P$ symmetry hypothesis.
  Studies of triple-product asymmetries, which provide an additional probe of $C\!P$ violation, show no significant deviation from $C\!P$ symmetry.

\end{abstract}

\vspace*{2.0cm}

\begin{center}
Published in Science Bulletin 71 (2026) 547
\end{center}

\vspace{\fill}

{\footnotesize 
\centerline{\copyright~\papercopyright. \href{\paperlicenceurl}{\paperlicence}.}}
\vspace*{2mm}

\end{titlepage}


\newpage
\setcounter{page}{2}
\mbox{~}
%
%
%
%


\renewcommand{\thefootnote}{\arabic{footnote}}
\setcounter{footnote}{0}


\cleardoublepage


\pagestyle{plain} 
\setcounter{page}{1}
\pagenumbering{arabic}



\section{Introduction}
Measurements of \CP violation are a powerful tool for discovering physics beyond the Standard Model and addressing the long-standing matter-antimatter asymmetry puzzle\mbox{~\cite{Sakharov:1967dj,Dine:2003ax}}. Since the first observation of \CP violation in neutral kaons in 1964~\cite{PhysRevLett.13.138}, \CP violation has also been established in both the beauty and charm quark sectors~\cite{LHCb-PAPER-2012-038}. In $b$ hadrons, \CP violation  was first established in 2001 through the interference of decay and $\Bz$--$\Bzb$ mixing, although no direct \CP violation in decay was found~\cite{BaBar:2001pki}. Direct \CP violation was later observed in $B$-meson decays, \ie $\Bz\to\Kp \pim$ via $b\to s \uubar$ transitions~\cite{Belle:2004nch,BaBar:2004gyj}, where the interference between tree-level and so-called electroweak loop (penguin) diagrams plays a critical role.\footnote{Unless otherwise stated, the inclusion of charge conjugate processes is implied throughout the text.} Recently, the \lhcb experiment reported the first evidence of direct \CP asymmetry (\ACP) in the difference between $\Bp \to \jpsi \pip$ and $\Bp \to \jpsi \Kp$ decays,  governed by $b \to \ccbar q \,(q=d,s)$ transitions~\cite{LHCb-PAPER-2024-031}. 
The measurement of \mbox{\decay{\bquark}{\ccbar\dquark}} transitions plays a crucial role in constraining subleading penguin diagram contributions to \mbox{\decay{\bquark}{\ccbar\squark}} processes. 
This approach enables a cleaner extraction of the weak phase $2\beta\equiv 2\arg[-(V_{cd}^{}V_{cb}^{*})/(V_{td}^{}V_{tb}^{*})]$ through penguin-enhanced processes, thereby improving precision tests of the Standard Model~\cite{Fleischer:1999zi,Faller:2008gt,Dunietz:1993cg,Soares:1994vi,DeBruyn:2014oga,Barel:2020jvf,debruyn2025tamepenguinsadvancinghighprecision}. Here, $V_{ij}$ denotes an element of Cabibbo–Kobayashi–Maskawa (CKM) matrix, which describes the mixing of quark flavors under the weak interaction.

While evidence for $\CP$ violation has been found in beauty meson decays into charmonium~\cite{LHCb-PAPER-2024-031}, the same significance has not yet been established in the corresponding baryon sector~\cite{Gershon:2016fda}, 
despite similar quark-level dynamics and large predicted asymmetries~\cite{Dunietz:1993cg,Soares:1994vi}.
Exploring $\CP$ violation in the baryon sector provides additional insights into the mechanisms of $\CP$ violation in the Standard Model and may lead to the discovery of new physics~\cite{Dine:2003ax}.
The first observation of \CP violation in baryon decays has been achieved with the \mbox{\decay{\Lb}{\proton\Km\pip\pim}} decay by the \lhcb collaboration in 2025~\cite{LHCb-PAPER-2024-054}.
Recent measurements of \mbox{$\Lb \to \Lz K^- K^+$} decays suggest an evidence of $\CP$ violation with a significance of 3.1 standard deviations ($\sigma$), which requires confirmation with additional data~\cite{LHCb-PAPER-2024-043}. 
Asymmetries in baryon decays have also been explored in other decay channels, including  
\mbox{\decay{\Lb}{\Dz \proton\Km}}~\cite{LHCb-PAPER-2021-027},   \mbox{\decay{\Lb}{\proton\pim}}~\cite{LHCb-PAPER-2018-025}, \mbox{\decay{\Xibm}{\proton\Km\Km}}~\cite{LHCb-PAPER-2020-017}, 
\mbox{\decay{\Lb}{\proton\Km\mup\mun}}~\cite{LHCb-PAPER-2016-059}, 
\mbox{\decay{\Lb}{\KS\proton\pim}}~\cite{LHCb-PAPER-2013-061}, 
\mbox{\decay{\Lb}{\proton \pim \pip \pim}}~\cite{LHCb-PAPER-2019-028, LHCb-PAPER-2016-030} and \ensuremath{\mbox{\decay{\Lb/\Xibz}{\proton h h^{'} h^{''}}}}~\cite{LHCb-PAPER-2018-044, LHCb-PAPER-2018-001}, where $h$ represents a kaon or pion. No  \CP violation has been observed in these decay modes to date.

Decays of the $\Lb$ baryon into charmonium final states, such as \LbJpsipK and \LbJpsippi, provide a baryonic counterpart to the mesonic $b \to \ccbar q$ transitions discussed above.
The \LbJpsipK decay, dominated by the $b \to \ccbar s$ transition, is expected to receive negligible penguin contributions and thus provides a reference channel with minimal \CP violation, similar to the role of $b \to \ccbar s$ mesonic modes.
In contrast, the \LbJpsippi decay proceeds via a $b \to \ccbar d$ transition, where penguin amplitudes are enhanced, making it sensitive to potentially larger \CP asymmetries, mirroring the situation in mesonic $b \to \ccbar d$ case. 
The interference pattern can be expressed in terms of tree and penguin amplitudes as $\mathcal{M} \sim (V_{cd} V_{cb}^\ast)T + (V_{td} V_{tb}^\ast)P_t$ for \LbJpsippi and $\mathcal{M} \sim (V_{cs} V_{cb}^\ast)T + (V_{us} V_{ub}^\ast)P_u$ for \LbJpsipK, where $T$ and $P$ denote tree-level and penguin contributions, respectively~\cite{PDG2024}.
Measuring the \CP asymmetry difference between these two decay modes, 
\begin{equation}
\label{eq:def_dacp}
    \DACP \equiv \ACP(\decay{\Lb}{\jpsi\proton\pim}) - \ACP(\decay{\Lb}{\jpsi\proton\Km}),
\end{equation} 
provides crucial information not only for constraining the penguin-induced phase shift but also for testing the Standard Model's predictions in baryonic decays.

The Cabibbo-favored decay \LbJpsipK was first observed in 2013 by the \lhcb collaboration and was used for a precise lifetime measurement of the $\Lb$ baryon~\cite{LHCb-PAPER-2013-032}, based on proton-proton ($pp$) collision data collected with the \lhcb detector at center-of-mass energies of 7 and 8\tev during Run~1 (2011--2012), corresponding to an integrated luminosity of 3\invfb. Using the same dataset, the Cabibbo-suppressed decay \LbJpsippi was first observed in 2014. This analysis also provided a measurement of \DACP between the two decays, yielding $(5.7\pm2.4\pm1.2)\%$~\cite{LHCb-PAPER-2014-020}, where the first uncertainty is statistical and the second is systematic. 
This article presents an updated measurement of \DACP using data recorded at 13\tev during Run~2 (2015--2018), corresponding to an integrated luminosity of 6\invfb. The results are obtained by integrating over the full phase space of the $\Lb$ decay, defined in terms of the \jpsi\proton and \proton\pim invariant masses, without accounting for the variation of the experimental efficiency across the phase-space regions.

\section{The \lhcb Detector}
The LHCb 
detector~\cite{LHCb-DP-2008-001,LHCb-DP-2014-002} is a
single-arm forward spectrometer covering the pseudorapidity range $2 < \eta < 5$, designed for
the study of particles containing \bquark\ or \cquark\ quarks. The detector used for this analysis includes a high precision tracking system consisting of a silicon-strip vertex detector surrounding the $pp$ interaction
region, a large-area silicon-strip detector located upstream of a dipole magnet with a bending power of about 4\,T\,m, and three stations of silicon-strip detectors and straw drift tubes placed downstream of the magnet.
The tracking system provides a measurement of the momentum, $p$, of charged particles with a relative uncertainty ranging from 0.5\% at low momentum to 1.0\% at 200\gevc, and determines the impact parameter with a resolution of $(15 + 29 / \pt)\mum$, where \pt is the transverse momentum of the particle in\,\gevc.
Particle identification (PID) of final-state particles is achieved using two ring-imaging Cherenkov detectors, a calorimeter system, and a muon detection system.

Simulated samples of the decays \LbJpsippi and \LbJpsipK are generated according to a uniform distribution in three-body phase space, 
to optimize the event selection, to
constrain the fit models for the determination of the signal (reference) yields, and to describe the shape of misidentified backgrounds. 
In the simulation, $pp$ collisions are generated using \pythia~\cite{Sjostrand:2007gs,*Sjostrand:2006za} with a specific \lhcb configuration~\cite{LHCb-PROC-2010-056}.
Decays of unstable particles are described by \evtgen~\cite{Lange:2001uf}, in which final-state radiation is generated using \photos~\cite{davidson2015photos}.
The interaction of the generated particles with the detector, and its response,
are implemented using the \geant toolkit~\cite{Allison:2006ve, *Agostinelli:2002hh} as described in Ref.~\cite{LHCb-PROC-2011-006}.
The kinematics of the \Lb baryon and the PID of final-state particles in the simulated samples are corrected to account for the discrepancies between simulation and data.

\section{Candidate selection}

The online event selection is performed using a trigger system~\cite{LHCb-DP-2012-004}, which consists of a hardware stage followed by a two-level software stage designed to select \mbox{$\jpsi\to \mup \mun$} candidates. The hardware stage relies on information from the calorimeter and muon systems and requires events to contain at least one muon with high \pt. The software stage operates in two parts: the first stage performs a partial reconstruction, requiring either a pair of well-reconstructed, oppositely charged muons with an invariant mass greater than 2.7\gevcc, or a single well-reconstructed muon with high \pt and a significant impact parameter with respect to any primary $pp$ collision vertex (PV). The second stage ensures that the pair of oppositely charged muons forms a good vertex-fit chisquare that is well separated from any PV.

In the offline analysis, additional requirements are applied on candidates selected by the trigger. The $\jpsi$ candidates are reconstructed from a $\mup\mun$ pair, where each muon has $\pt > 500\mevc$. 
The invariant mass of the muon pair is required to lie within the range [3049, 3140]\,\mevcc.
The $\Lb$ candidates are formed by combining the $\jpsi$ candidate with a proton and a negatively charged hadron $h^-$, where $h^-$ denotes either $\pi^-$ or $K^-$. The hadrons (\proton, $h^-$) are required to have a good track-fit chisquare, high \pt, positive particle identification, and must be inconsistent with originating from any PV. The $\jpsi$, proton, and $h^-$ must originate from a common vertex (\Lb vertex). The $\Lb$ candidate is required to be consistent with originating from a PV, and to have a flight distance greater than 1.5\,mm. To distinguish \LbJpsippi and \LbJpsipK decays, the information from the tracking system and Cherenkov detectors is used to form a probability for the hypothesis of a final-state particle being a pion or kaon, calculated by a neural network~\cite{LHCb-DP-2012-003}. The pion candidate probability of being a pion should be greater than that of being a kaon, and vice versa for the kaon candidate.
In addition, fiducial-volume requirements are imposed to exclude $\Lb$ candidates whose decay products are near the boundaries of the detector acceptance, where detection asymmetry is significant.

For the signal \LbJpsippi channel, backgrounds due to particle misidentification from \mbox{\decay{\Bs}{\jpsi \pip \pim}}, \mbox{\decay{\Bs}{\jpsi \Kp \Km}}, \mbox{\decay{\Lb}{\jpsi \proton \Km}}, and \mbox{\decay{\Lbbar}{\jpsi \Kp \antiproton}} decays are removed by applying tighter PID requirements on the decay products whose invariant mass lies within a $\pm 22$\mevcc window (twice the mass resolution) around the nominal mass of the corresponding $b$ hadron~\cite{PDG2024}. 
Another source of background arises from \mbox{\decay{\Bd}{\jpsi \Kp \pim}} decays. To suppress such background, candidates are rejected if the invariant mass $m(\jpsi \Kp \pim)$, calculated by assigning the kaon mass hypothesis to the proton candidate, falls within a mass window of $\pm 33\mevcc$ around the known \Bd mass~\cite{PDG2024}, and either the $\Kp \pim$ invariant mass lies within $\pm 100$\mevcc of the known $K^*(892)^0$ mass~\cite{PDG2024}, or the proton candidate has a high probability of being misidentified as a kaon. With the above selections, there still remain contributions from \mbox{\decay{\Lb}{\jpsi \proton \Km}} decays which are considered as a background component in the fit used to determine the signal yields. 
Furthermore, residual background from \mbox{\decay{\Lb}{\jpsi\Lz}}, with 
long-lived \mbox{\decay{\Lz}{\proton\pim}} decays is excluded by rejecting candidates with the $p\pim$ invariant mass within $\pm4\mevcc$ of the known $\Lz$ mass~\cite{PDG2024}. For the reference \LbJpsipK channel, a similar procedure is applied to remove the \mbox{\decay{\Bs}{\jpsi \Kp \Km}}, \mbox{\decay{\Bzb}{\jpsi \Kp \Km}}, and \mbox{\decay{\Lbbar}{\jpsi \Kp \antiproton}} contributions. These selection requirements are the same as those used in Ref.~\cite{LHCb-PAPER-2019-014}.

Combinatorial background is  suppressed using a multivariate analysis based on a gradient boosted decision tree (BDTG) classifier implemented in the TMVA toolkit~\cite{Friedman:2001wbq,Hocker:2007ht}.
The discriminant variables comprise kinematic properties, track and vertex reconstruction quality, and particle identification information for protons and muons.
The BDTG is trained on a simulated sample for the signal and data candidates in the mass range \mbox{$m(\jpsi\proton\pim) \in [5770, 5870]\mevcc$} for the background. The requirement on the BDTG response is optimized by maximizing the signal significance $S/\sqrt{S+B}$, where $S$ and $B$ stand for signal and background yields in the signal region, defined as \mbox{$m(\jpsi\proton\pim) \in [5595, 5645]\mevcc$}. The BDTG response optimization retains 73\% of signal candidates and rejects 98\% background candidates for the \LbJpsippi decays.  The same BDTG response and working point are applied to the reference channel.

\section{Invariant-mass fits}
The yields of signal and reference channels are extracted using an extended unbinned maximum-likelihood fit, performed simultaneously on the \Lb and \Lbbar mass distributions. 
The $\Lb$ signal shape is modeled by a Hypatia function~\cite{Santos:2013gra}. Tail parameters for the signal and reference channels are derived from their corresponding simulated samples and fixed during the fit to data, while the Hypatia width and peak position are free parameters to accommodate discrepancies between simulation and data.
The combinatorial background is modeled by an exponential function. The residual background from \LbJpsipK decays to the signal sample is modeled using simulation, where the kaon mass is assigned to the pion to emulate its contribution to the \jpsi\proton\pim invariant-mass spectrum.
The parameters of the probability density functions are shared between the $\Lb$ decay and its charge conjugate, except for the yields of each component.
The fits are performed separately for the \LbJpsippi and \LbJpsipK decays.

Nuisance asymmetries arising from the production of $\Lb$ and $\Lbbar$ baryons, as well as from proton detection efficiency, are accounted for in the simultaneous fits.
These asymmetries arise from the different production cross-sections of \Lb and \Lbbar baryons in $pp$ collisions, which depend on the $\eta$ and \pt of \Lb baryons. Additionally, differences in the interaction of protons and antiprotons with matter lead to distinct detection efficiencies~\cite{LHCb-PAPER-2018-025}. 
To cancel these asymmetries, the five-dimensional distribution of $(p, \pt, \eta)$ of the \Lb baryons and the $(p, \pt)$ of the proton of the \LbJpsipK decay, is weighted to match that of the \LbJpsippi decay. 
This weighting is achieved using the \sPlot method~\cite{Pivk:2004ty} to statistically subtract the background and the \texttt{hep\underline{~}ml} package~\cite{Rogozhnikov:2016bdp} to calculate the weights.
The log-likelihood of the $\Lb$ mass fit is scaled by a factor that takes into account the impact of the weighting procedure on the statistical uncertainty, using Kish's effective sample size method~\cite{kish1965survey, Xie:2009rka,LHCb-PAPER-2013-002, Langenbruch:2019nwe}. 

\begin{figure}[!t]
  \begin{center}
    \begin{subfigure}{0.4\linewidth}
    \centering
    \includegraphics[width=\linewidth]{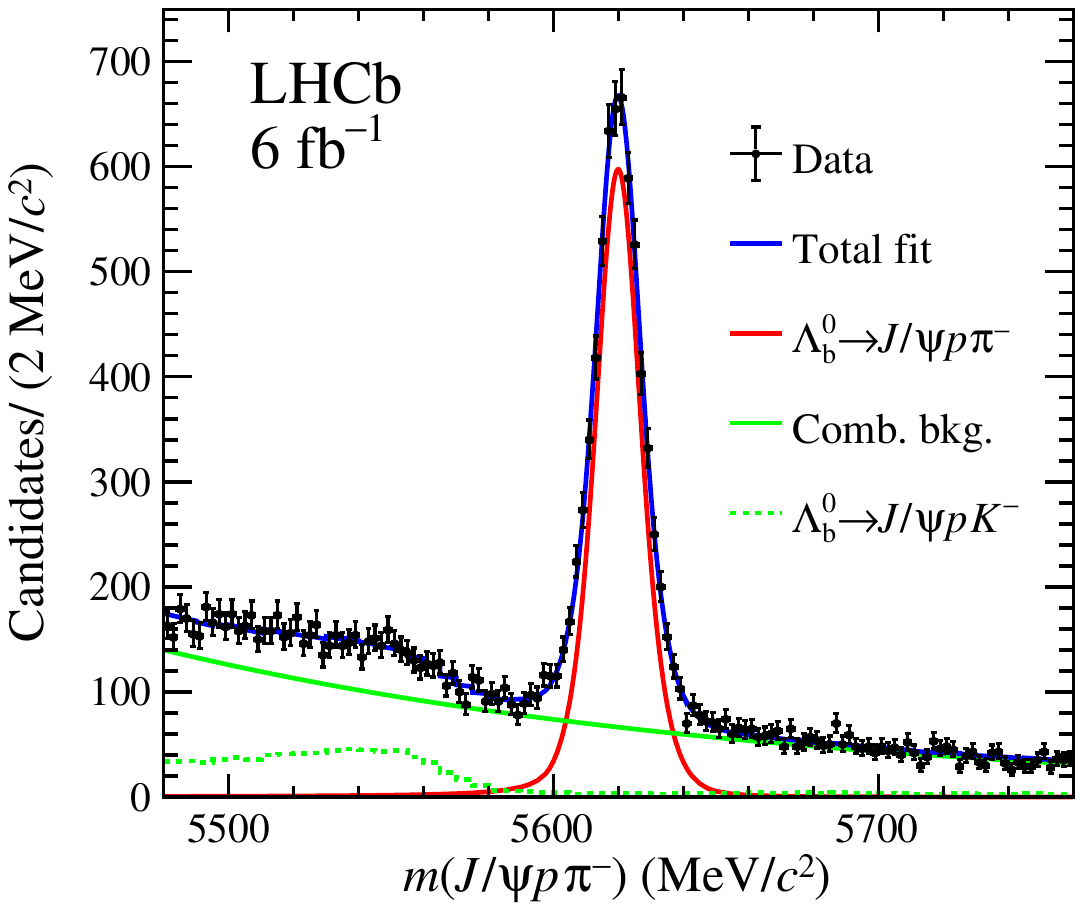}
    \caption{}
    \end{subfigure}
    \begin{subfigure}{0.4\linewidth}
    \centering
    \includegraphics[width=\linewidth]{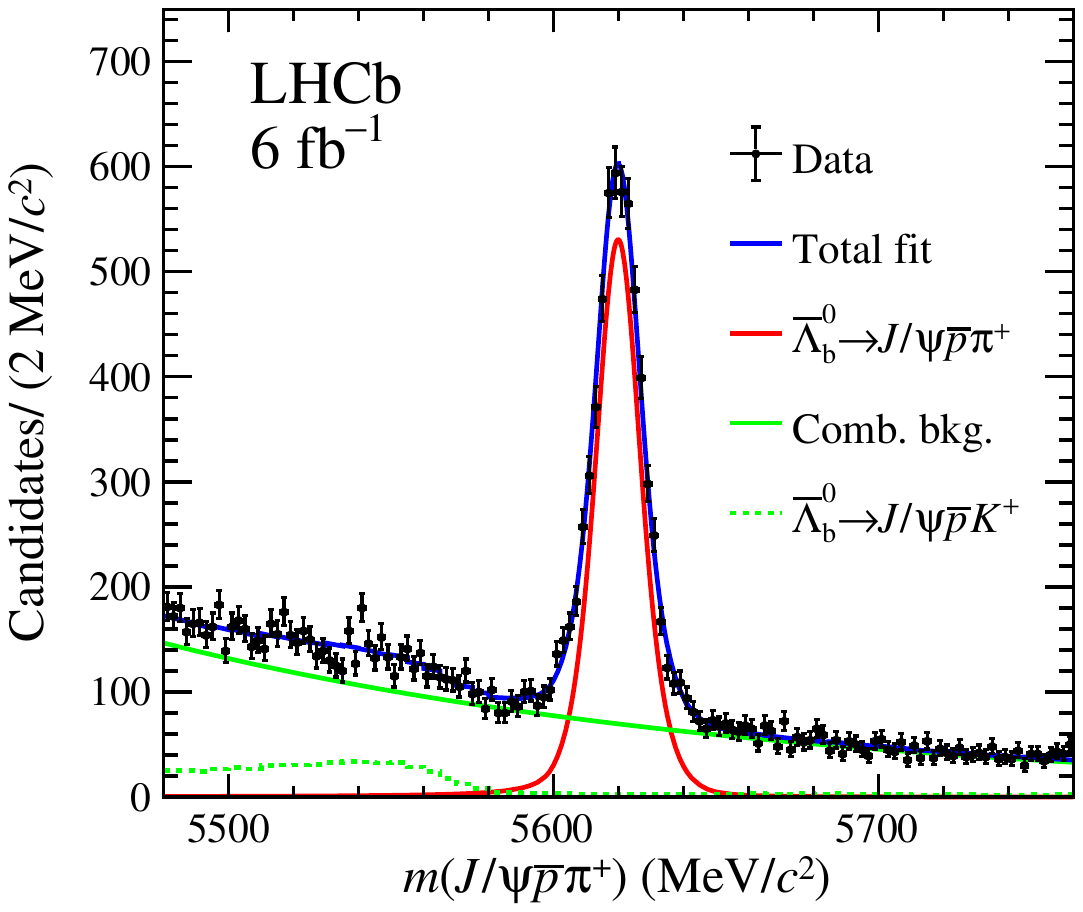}
    \caption{}
    \end{subfigure}
    \vspace*{-0.5cm}
  \end{center}
  \caption{
    \small 
    Mass distributions of (a) ${\Lambda}^0_b \!\rightarrow J/\psi p \pi^-$ and (b) ${\overline{\Lambda}}{}^0_b \!\rightarrow J/\psi \overline{p} \pi^+$ candidates with the results of the fit also shown. }
  \label{fig:acp-pi}
\end{figure}

\begin{figure}[!t]
  \begin{center}
    \begin{subfigure}{0.4\linewidth}
    \centering
    \includegraphics[width=\linewidth]{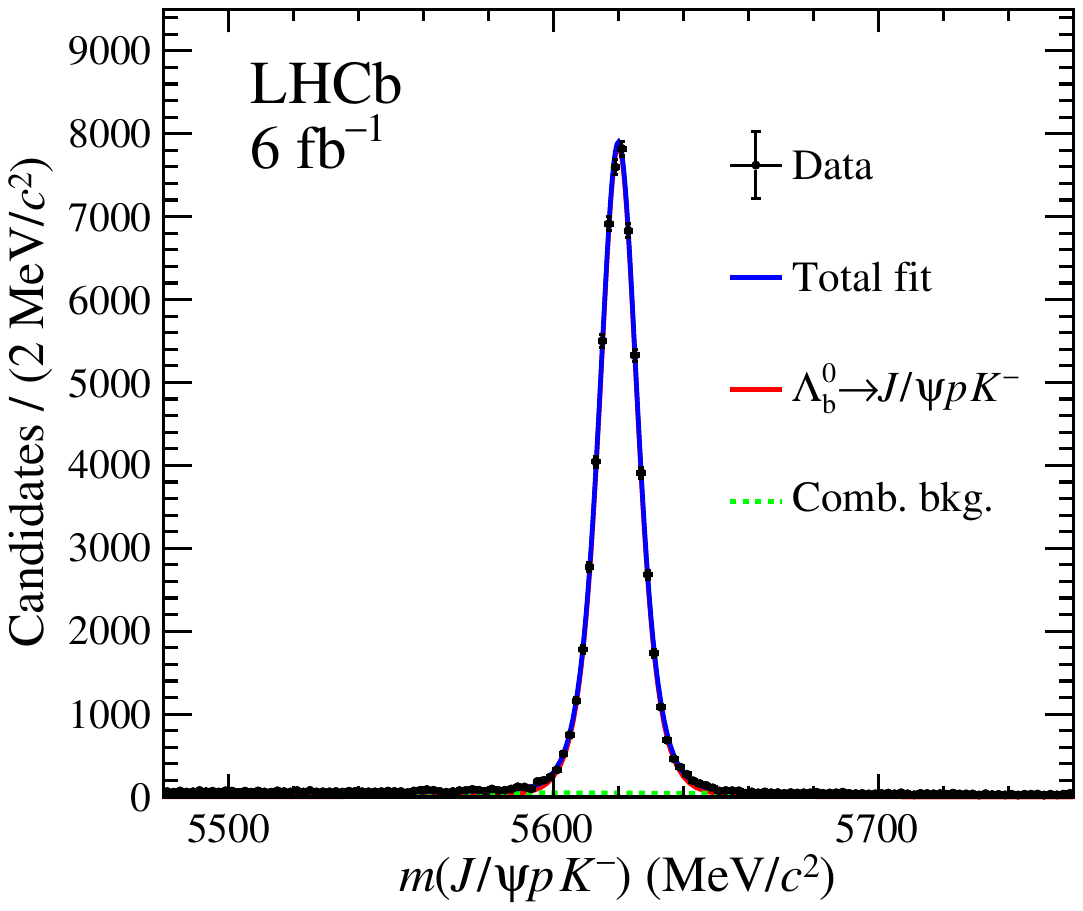}
    \caption{}
    \end{subfigure}
    \begin{subfigure}{0.4\linewidth}
    \centering
    \includegraphics[width=\linewidth]{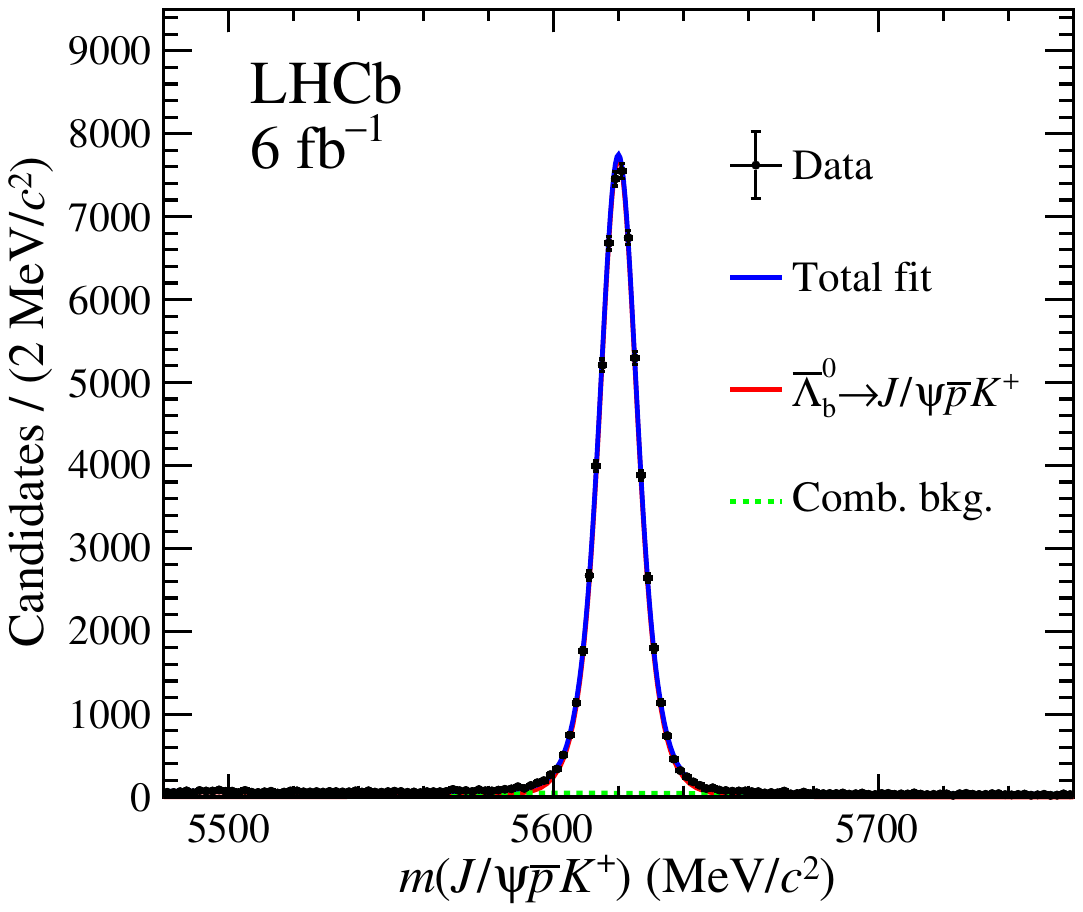}
    \caption{}
    \end{subfigure}
    \vspace*{-0.5cm}
  \end{center}
  \caption{
    \small 
    Mass distributions of (a) ${\Lambda}^0_b \!\rightarrow J/\psi p K^-$ and (b) ${\overline{\Lambda}}{}^0_b \!\rightarrow J/\psi \overline{p} K^+$ candidates with the results of the fit also shown. }
  \label{fig:acp-k}
\end{figure}

The measured yields are $N(\LbJpsippi) = 10853\pm134$ and $N(\LbJpsipK) = 125380\pm372$, respectively. The signal yield for $\LbJpsippi$ is approximately 5.7 times larger than that obtained in our previous analysis~\cite{LHCb-PAPER-2016-015}.
The ``raw'' asymmetry between $\Lb$ and $\Lbbar$ decays, denoted as \Araw, is determined as the relative difference in their yields obtained from the invariant-mass fits, as shown in Figs.~\ref{fig:acp-pi} and~\ref{fig:acp-k}.
The fit results are $\Araw (\LbJpsippi) = (\varrpiraw\pm\varrpirawerr)\%$ and $\Araw(\LbJpsipK) = (\varrkraw)\%$.
The detection asymmetries of pions and kaons differ and do not cancel in the calculation of $\DACP$, requiring a correction to the raw $\DACP$ determined from their difference.
The kaon and pion detection asymmetries are functions of their momenta, and can be determined from their kinematic distributions.
At LHCb, the difference in raw asymmetries of the \mbox{\decay{\Dp}{\KS\pip}} and \mbox{\decay{\Dp}{\Km\pip\pip}} decays can probe the two-track detection asymmetry ${\cal A}_\mathrm{D}(\Km) - {\cal A}_\mathrm{D}(\pim)$. 
Using the LHCb calibration data collected during 2016--2018, following the methods outlined in Refs.~\cite{LHCb-PUB-2018-004,LHCb-PAPER-2014-069,LHCb-PAPER-2016-054,LHCb-PAPER-2024-048}, 
the detection asymmetry is estimated by
fitting the LHCb calibration dataset of these two $\Dp$ decays, with their kaon and pion kinematic distributions weighted to match the the background-subtracted distributions in this analysis.
The difference is estimated to be ${\cal A}_\mathrm{D}(\Km) - {\cal A}_\mathrm{D}(\pim) = (\varrad \pm \varraderr)\%$. Using this correction and the definition in Eq.~(\ref{eq:def_dacp}), $\DACP$ is calculated as 
\begin{equation}
\DACP = \Araw (\LbJpsippi) - \Araw (\LbJpsipK) + {\cal A}_\mathrm{D}(\Km) - {\cal A}_\mathrm{D}(\pim),
\end{equation}
and found to be
\begin{equation*}
    \DACP = (\varrtot \pm \varrtoterr) \%,
\end{equation*}
where the uncertainty is statistical, and is calculated solely by adding the two \Araw statistical uncertainties in quadrature, with the statistical uncertainty of ${\cal A}_\mathrm{D}(\Km) - {\cal A}_\mathrm{D}(\pim)$ propagated to the systematic uncertainty.


\section{Systematic uncertainty}
Various sources of systematic uncertainties on the \DACP measurement are considered, as listed in Table~\ref{tab:sys}.

To account for potential biases from using the same peak position and width parameters in the simultaneous fits, these parameters are allowed to vary independently for the $\Lb$ and $\Lbbar$ samples. 
Fits are performed by allowing the peak position only, the Hypatia width only, and both parameters, to vary independently instead of sharing the same values. The largest uncertainty is obtained when varying both. The uncertainties in the two decay channels are added in quadrature.

The systematic uncertainties related to 
the signal and the background shapes are considered by altering the fit models.
The signal shape is varied by using the sum of two Crystal Ball functions~\cite{Skwarnicki:1986xj} for both decay modes. Pseudoexperiments are generated based on the baseline fit results. Fits are performed to each pseudoexperiment using both the baseline model and the alternative signal shape. The mean and standard deviation of the differences between the two sets of fits are combined in quadrature, and the result is assigned as a systematic uncertainty.
The same method is used to determine the uncertainty on the combinatorial background shape, by replacing the default model with a Chebyshev polynomial.
Residual background from misidentified \mbox{\decay{\Bz}{\jpsi\Kp\pim}} decays in the signal channel is included as a source of systematic uncertainty. This component is modeled as an exponential function in the $\jpsi\proton\pim$ invariant-mass fit, with its shape parameter fixed from simulation and its yield %
fixed from a dedicated fit under an alternative hypothesis where the proton is assigned the kaon mass.

The uncertainty from weighting is evaluated by varying the procedure applied to the \LbJpsipK sample, using alternative schemes based on \sPlot results obtained from segmented phase-space %
regions of the \LbJpsippi decay.
The uncertainty from the efficiency asymmetry due to the PID requirements on the final-state hadrons is determined from calibration samples, and is calculated as the quadrature sum of the asymmetry's central value and uncertainty. The statistical uncertainty on the detection asymmetry \mbox{${\cal A}_\mathrm{D} (\Km) - \mathcal{A}_\mathrm{D}(\pim)$} is also incorporated into the systematic uncertainty.

Compared to the previous results~\cite{LHCb-PAPER-2014-020}, systematic uncertainties are significantly reduced owing to a much larger calibration sample for detection asymmetries and more optimized selection criteria that minimize background, thereby reducing the associated modeling uncertainties.

\begin{table}[!t]
  \caption{
    \small 
    Summary of absolute systematic uncertainties of $\Delta {\cal A}_{C\!P}$. The second column is the uncertainty of $\Delta {\cal A}_{C\!P}$. The third column is the uncertainty of the triple-product asymmetry $\mathcal{A}_{\widehat{T}-\mathrm{odd}}$ of the ${\Lambda}^0_b \!\rightarrow J/\psi p \pi^-$ decays, which is defined in Sect.~\ref{sec:tpa}.
    }
\begin{center}\begin{tabular}{l l l}
    \hline
    Source of uncertainty                         &{\DACP  (\%)} & \Atodd (\%)\\ \hline
    {\Lb peak parameters} & ${0.210}$ & \phantom{0}${0.210}$ \\
    {Signal shape} & ${\varrsigsys}$ & \phantom{0}${\varrsigsys}$\\
    {Background shape} & ${\varrbkgsys}$ & \phantom{0}${\varrbkgsys}$ \\
    {\decay{\Bd}{\jpsi\Kp\pim} background} & ${0.013}$ & \phantom{0}${0.013}$\\
    Weighting & 0.013 &\phantom{00}--- \\
    PID asymmetry & 0.074 & \phantom{00}---\\
    Detection asymmetry  & 0.040 & \phantom{00}--- \\
    \hline
    Total & 0.230 & \phantom{0}0.210 \\
    \hline
  \end{tabular}\end{center}
\label{tab:sys}
\end{table}
\section{Triple-product asymmetry}
\label{sec:tpa}
The direct \CP asymmetry is linked to the \CP-even phase difference, $\Delta \delta$, and the \CP-odd phase difference, $\Delta \phi$, between the tree-level and penguin diagrams, and is proportional to $\sin{\Delta\delta} \sin{\Delta\phi}$. 
If the decay amplitude is decomposed into $\widehat{T}$-even and $\widehat{T}$-odd terms, where $\widehat{T}$ denotes time reversal, the \CP-even phase can also be different between terms under this decomposition.
The triple-product asymmetry (TPA), which is proportional to $\cos{\Delta\delta'} \sin{\Delta\phi'}$, is investigated, where $\Delta\phi'$ ($\Delta \delta'$) denotes the relative weak (strong) phase between the $\widehat{T}$-odd and $\widehat{T}$-even part of the decay amplitudes~\cite{Durieux:2015zwa}. 
The scalar triple products of the final-state particle momenta in the $\Lb$ rest frame are defined as
\begin{equation}
  C_{\widehat{T}} \equiv \vec{p}_\mup \cdot ( \vec{p}_\proton \times \vec{p}_\pim)
\end{equation}
for $\Lb$ candidates, and 
\begin{equation}
  \overline{C}_{\widehat{T}} \equiv \vec{p}_\mun \cdot ( \vec{p}_\antiproton \times \vec{p}_\pip)
\end{equation}
for \Lbbar candidates, which change sign under time-reversal transformation.
The triple-product asymmetry is defined as the difference of yield asymmetries in \Lb and \Lbbar samples, as
\begin{equation}
  \Atodd \equiv \frac12 \left( \frac{N(C_{\widehat{T}}>0)-N(C_{\widehat{T}}\leq0)}{N(C_{\widehat{T}}>0)+N(C_{\widehat{T}}\leq0)} - \frac{\overline{N}(-\overline{C}_{\widehat{T}}>0)-\overline{N}(-\overline{C}_{\widehat{T}}\leq0)}{\overline{N}(-\overline{C}_{\widehat{T}}>0)+\overline{N}(-\overline{C}_{\widehat{T}}\leq0)} \right).
  \label{eqn:tpa}
\end{equation}
The observable $\Atodd$ is particularly advantageous for experimental measurements, as any non-\CP asymmetry contributions related to production or detection are effectively canceled. 

The TPA in \LbJpsippi decays is measured by
performing separate fits to the \Lb and \Lbbar samples.  The \Lb samples with $C_{\widehat{T}}>0$ and $C_{\widehat{T}}\leq0$, as well as $-\overline{C}_{\widehat{T}}>0$ and $-\overline{C}_{\widehat{T}}\leq0$, are fitted simultaneously, with shape parameters shared for all components.
The fit model is the same as that used for measuring $\DACP$, except that no kinematic weight is applied to the $\jpsi p K^-$ sample, which is not relevant for this measurement.
The measured TPA is $\Atodd(\LbJpsippi) = (\varrpitpa)\%$, where the uncertainty is statistical only. The systematic uncertainty of $\Atodd(\LbJpsippi)$ is calculated similarly to that of \DACP, as listed in Table~\ref{tab:sys}.
\begin{figure}[!tb]
  \begin{center}
    \includegraphics[width=0.40\linewidth]{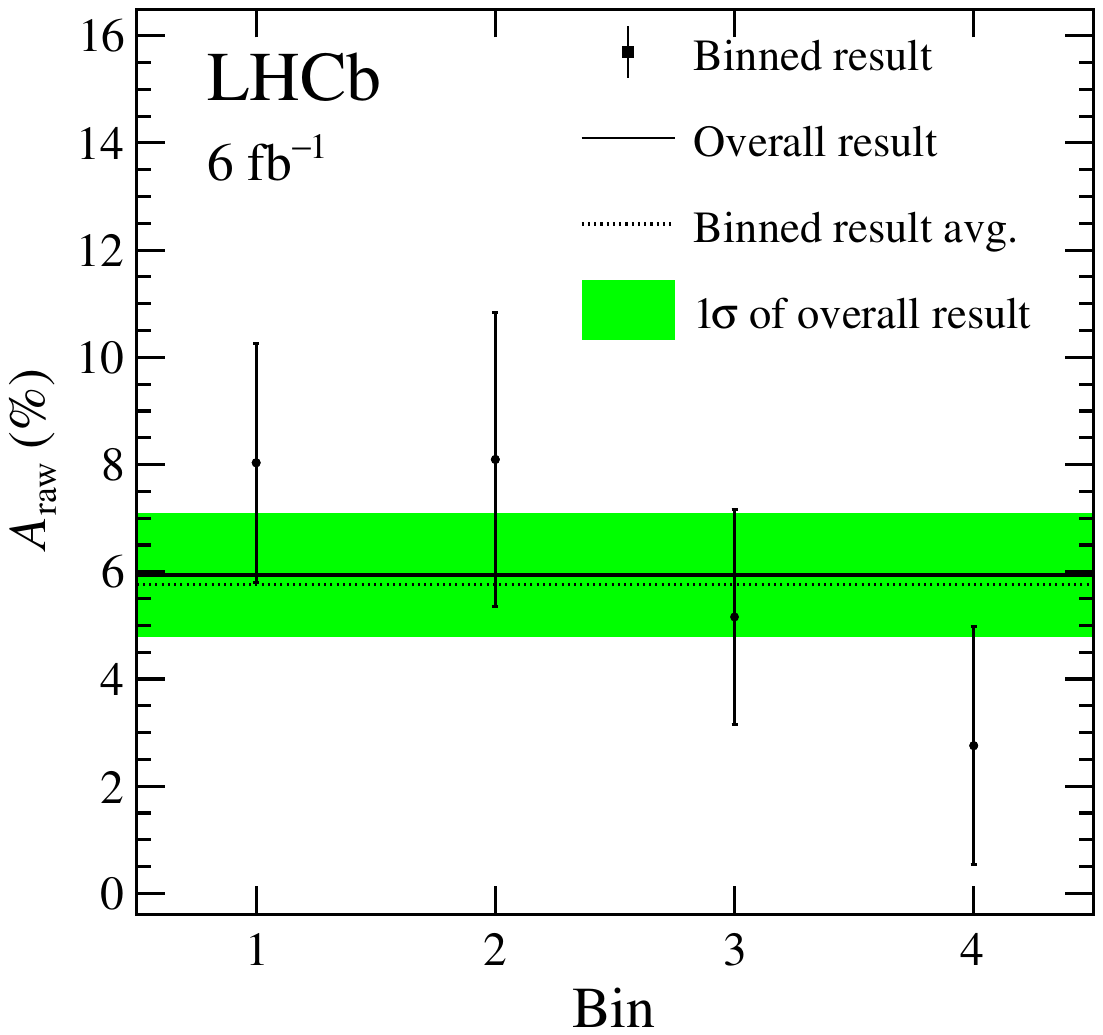}
    \includegraphics[width=0.40\linewidth]{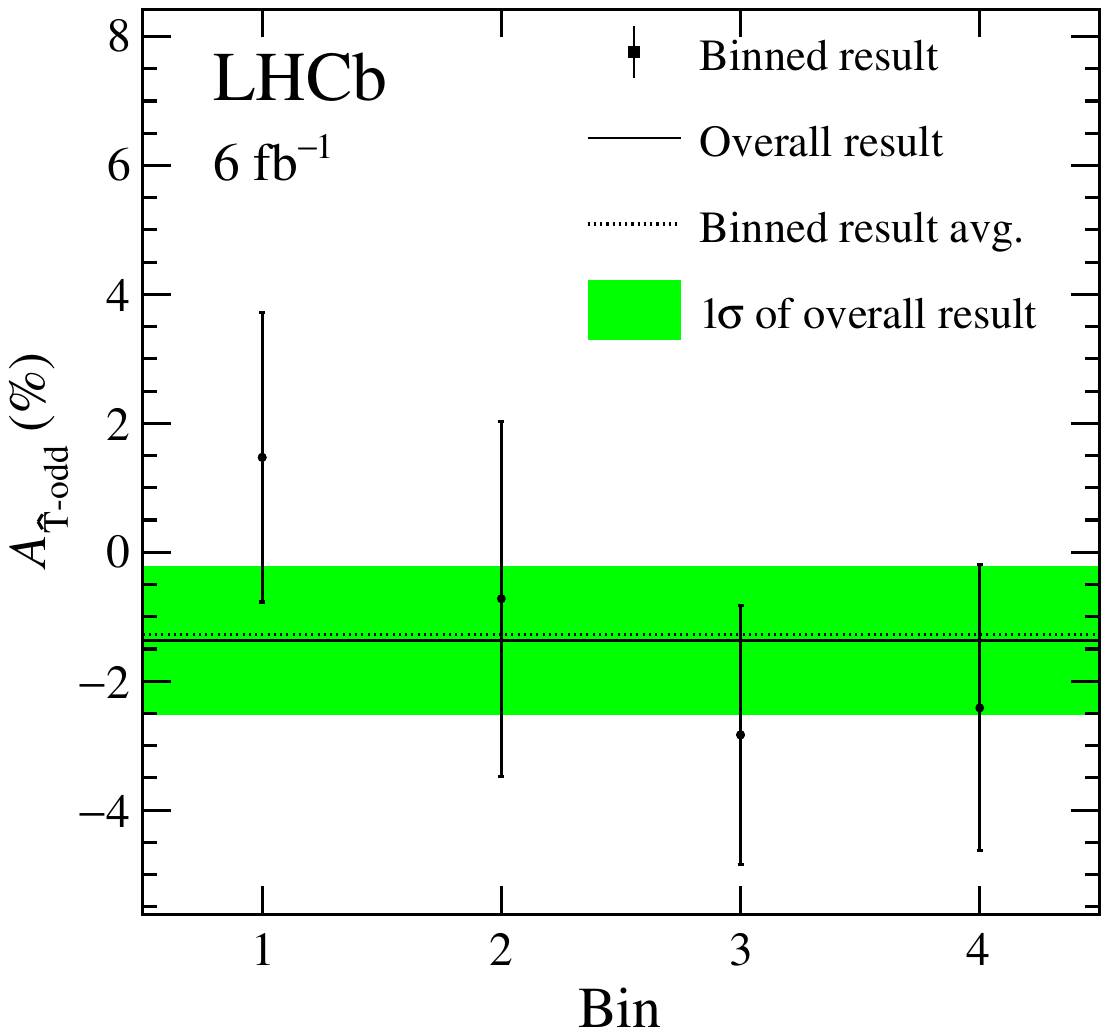}\\
    \includegraphics[width=0.40\linewidth]{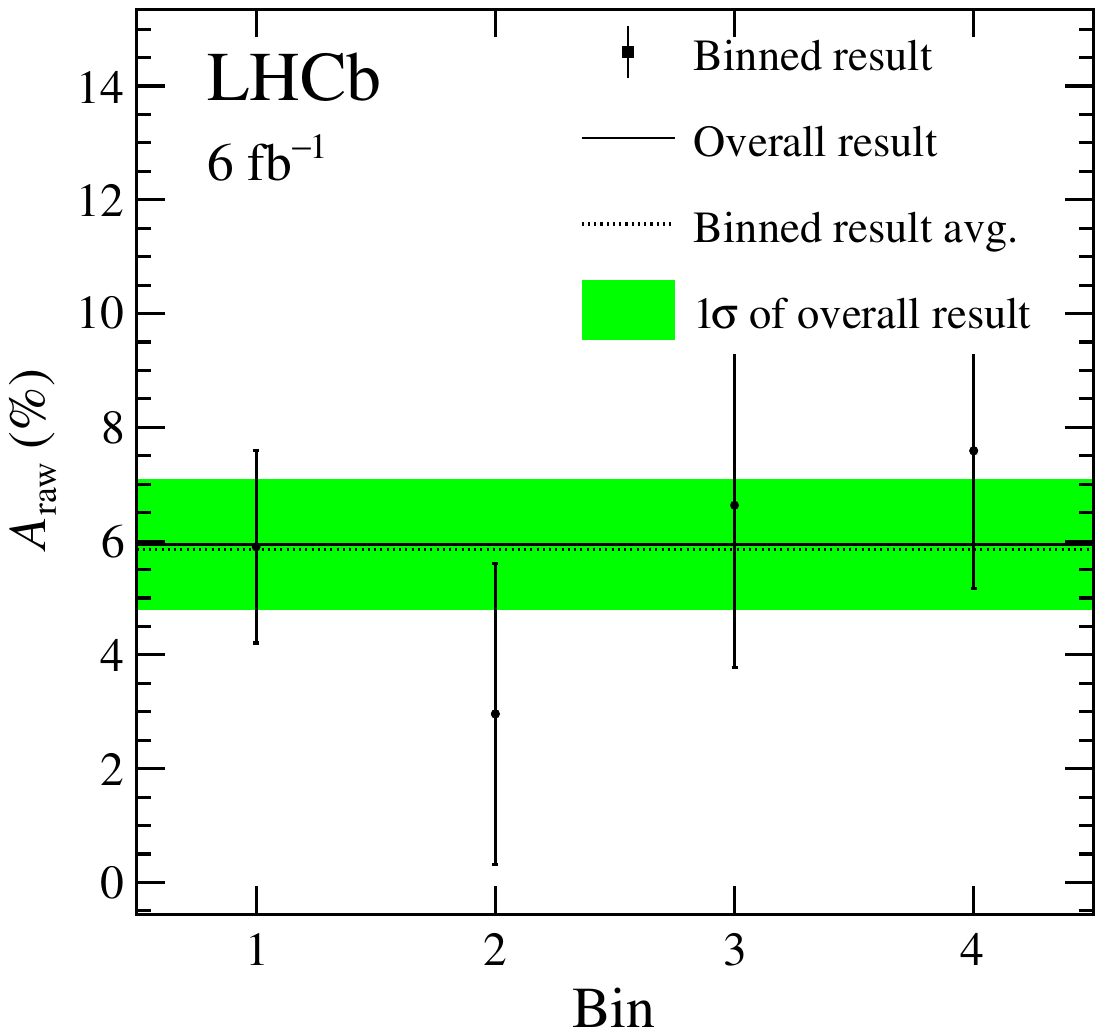}
    \includegraphics[width=0.40\linewidth]{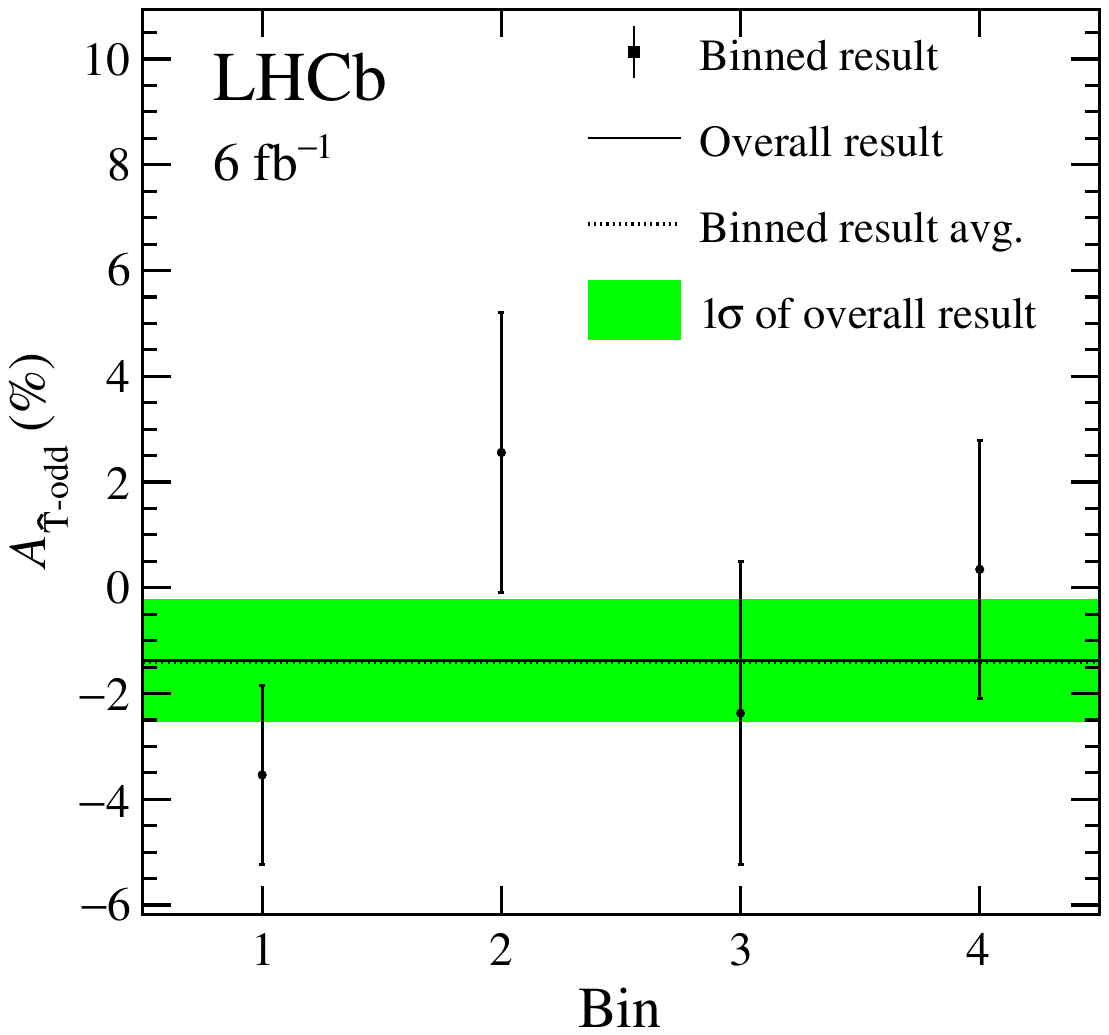}\\
    \includegraphics[width=0.40\linewidth]{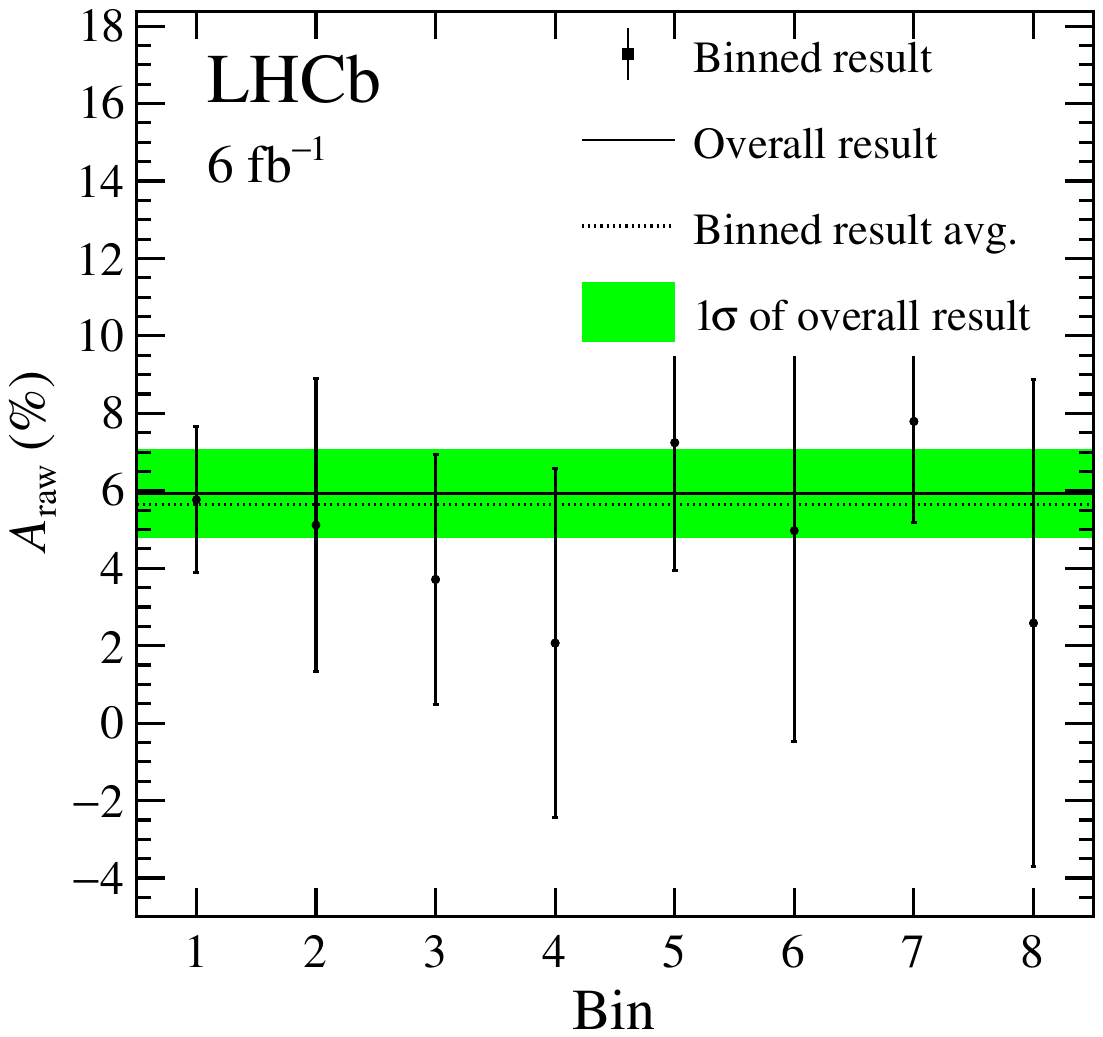}
    \includegraphics[width=0.40\linewidth]{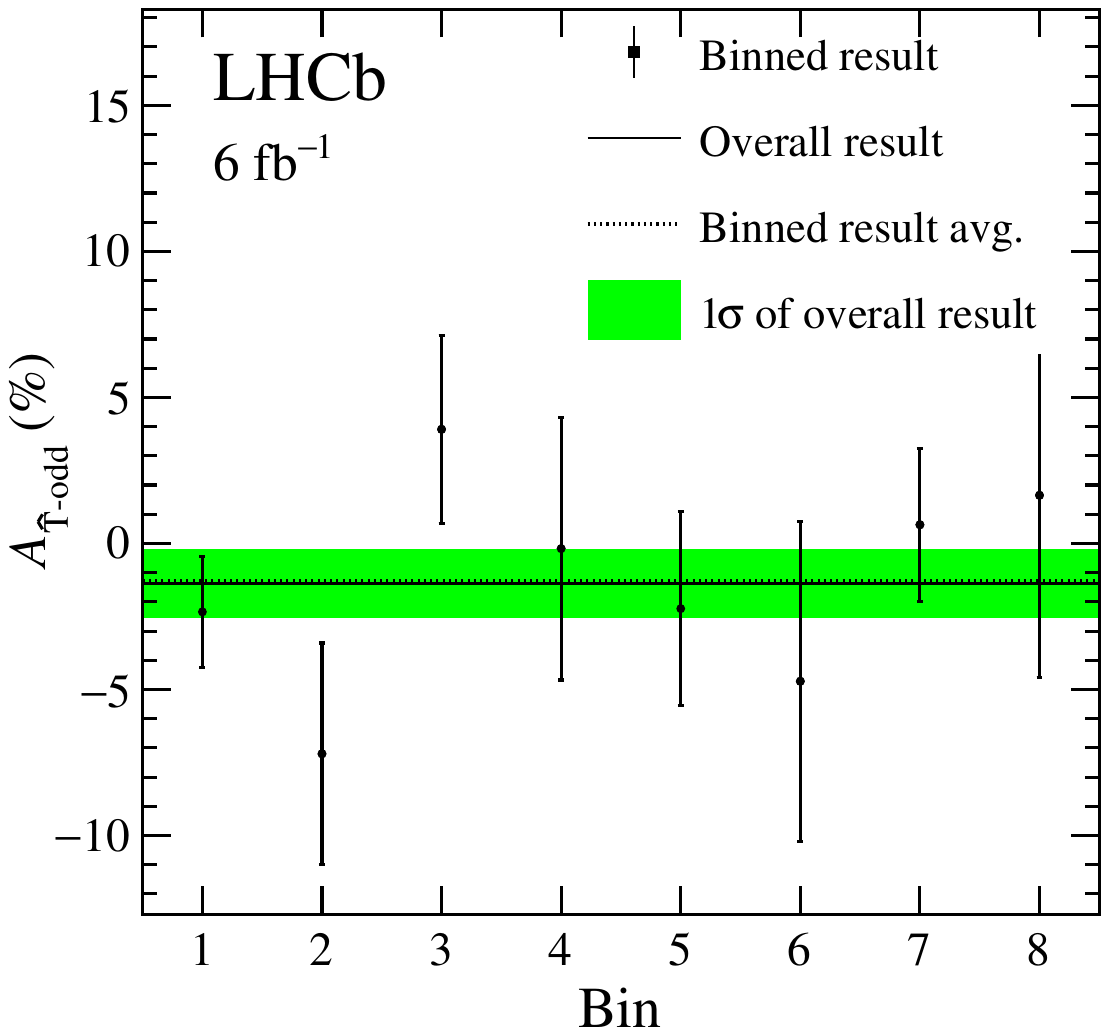}\\
    \vspace*{-0.5cm}
  \end{center}
  \caption{
    \small Results of the binned fit for the (left) raw asymmetry and (right) triple-product asymmetry. From top to bottom: 4 bins with equal candidate number; 4 bins separated by resonance peaks; $4\times2$ bins separated by resonant peaks and  sign of $\cos \theta$. }
  \label{fig:binned-1}
\end{figure}

\begin{figure}[!tb]
  \begin{center}
    \begin{subfigure}{0.48\linewidth}
    \centering
    \includegraphics[width=\linewidth]{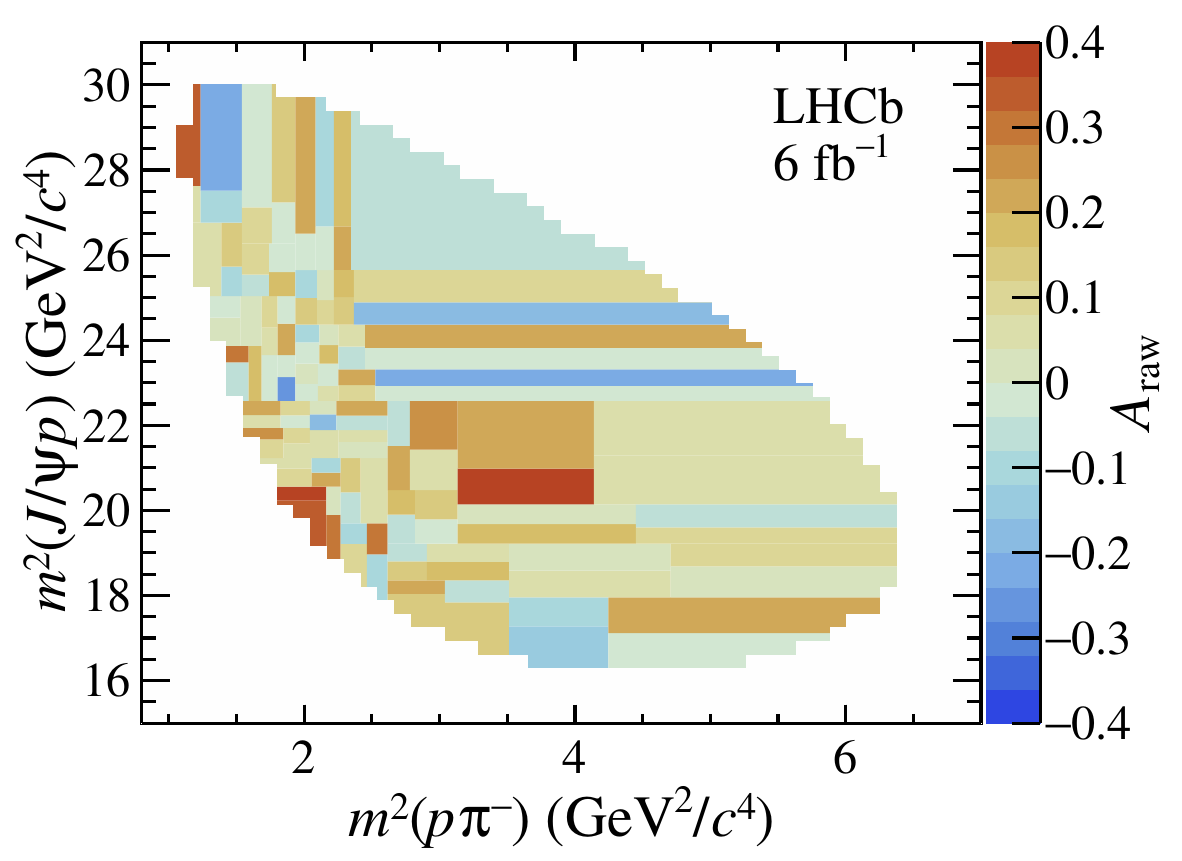}
    \caption{}
    \end{subfigure}
    \begin{subfigure}{0.48\linewidth}
    \centering
    \includegraphics[width=\linewidth]{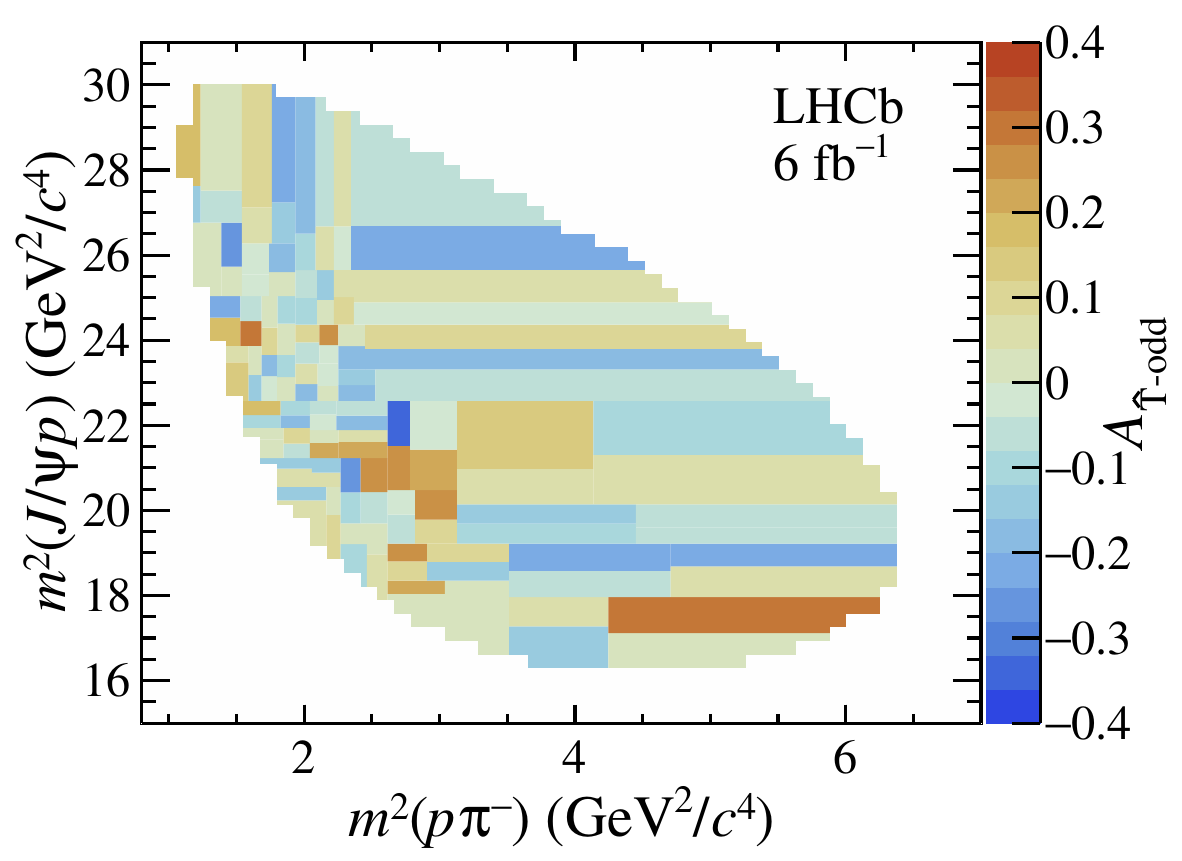}
    \caption{}
    \end{subfigure}
    \vspace*{-0.5cm}
  \end{center}
  \caption{
    \small The asymmetries in 128 bins. (a) Raw asymmetry; (b) triple-product asymmetry.%
    }
  \label{fig:binned-2}
\end{figure}

\section{Binned analyses}
Asymmetries in multibody decays often vary across the phase space of the final-state particles. To explore this, several binning schemes are employed. First, the phase space of the $\LbJpsippi$ decay is divided into 4 or 128 bins using an adaptive binning scheme. Each region contains an approximately equal number of candidates, to keep the uncertainty in different bins at the same level and facilitate the estimation of the significance of the observed variations. Second, the data are partitioned according to resonant structures to study the contributions of specific intermediate states to the global \CP asymmetry. The dominant contribution in this channel is from $\Lb \to \jpsi N (\to p\pim)$, where $N$ represents an excited nucleon resonance. Based on the $m(p\pim)$ distribution, the dataset is divided into four mass intervals: (1.07, 1.45], (1.45, 1.52], (1.52, 1.65], and (1.65, 3.00]\,\gevcc. Each of these four subsamples is further split according to the helicity angle $\theta$ of the $p\pim$ system, defined as the angle between the proton momentum and the opposite of the $\Lb$ momentum in the $p\pim$ rest frame. The splitting is done at $\cos{\theta} = 0$, yielding two bins per mass region.

Both the raw asymmetry $\Araw$ and TPA $\Atodd$ for the \LbJpsippi decay in these bins are measured,  and shown in Figs.~\ref{fig:binned-1} and~\ref{fig:binned-2}. The compatibility with the global result is evaluated by means of a $\chisq$ test, defined as $\chi^2 = R^T V^{-1} R$, where $R$ is the vector of differences between the bin-by-bin asymmetries and the corresponding global values, and $V$ is the corresponding statistical covariance matrix. 
The $\chi^2/\mathrm{ndof}$ (and corresponding $p$-values) for the adaptive binning scheme with 4 bins are 3.7/3 (0.30) for $\Araw$ and 2.4/3 (0.49) for $\Atodd$, while for 128 bins they are 120.5/127 (0.65) and 115.8/127 (0.75), respectively. 
Binning by $m(p\pim)$ yields $\chi^2/\mathrm{ndof}$ values of 1.8/3 (0.61) for $\Araw$ and 4.5/3 (0.21) for $\Atodd$. When further dividing according to the $\cos\theta$ value, the results are 2.2/7 (0.95) and 6.6/7 (0.47), respectively, indicating no significant variation of the asymmetries across the phase space.

\section{Conclusion}

In summary, a search for \CP violation in $\LbJpsippi$ decays is performed using $pp$ collisions collected by the \lhcb experiment during the 2015--2018 run period of the LHC. 
Both \CP and triple-product asymmetries have been measured to exploit the full potential of the data.
The phase-space integrated \CP asymmetry for the $\LbJpsippi$ decay, with respect to the $\LbJpsipK$ decay, is measured to be
\begin{equation*}
    \DACP \equiv \ACP (\LbJpsippi) - \ACP (\LbJpsipK) = (\varrtot \pm \varrtoterr \pm \varrresultsys)\%,
\end{equation*}
where the first uncertainty is statistical and the second is systematic, here and in the following.
The $\DACP$ measurement provides the first evidence of \CP violation in decays of \mbox{$b$ baryons} to charmonium, with a significance of $3.3\sigma$ when including systematic uncertainties.
This result is then combined with the \lhcb 2011--2012 result \mbox{$(5.7\pm2.4\pm1.2)\%$}~\cite{LHCb-PAPER-2014-020} with the Best Linear Unbiased Estimate (BLUE) method~\cite{Lyons:1988rp,Valassi:2003mu,Nisius:2020jmf}, obtaining 
\begin{equation*}
    \DACP = (\varrcomb \pm \varrcombstat  \pm \varrcombsys )\%.
\end{equation*}
The correlation in systematic uncertainties is considered by assuming a $100\%$ correlation in signal and combinatorial background shapes.
The combined result corresponds to a significance level of $3.9\sigma$.

The phase-space integrated triple-product asymmetry for the $\LbJpsippi$ decay is measured to be 
\begin{equation*}
\Atodd (\LbJpsippi) = (\varrresulttpa\pm \varrresulttpaerr  \pm \varrresulttpasys)\%,  
\end{equation*}
consistent with $\widehat{T}$-odd symmetry. 
The triple-product asymmetry \Atodd is sensitive to decay angular structures, notably the interference between different angular momenta~\cite{Gronau:2011cf,Durieux:2015zwa,Wang:2022fih,Wang:2024qff}, thereby complementing the angular-integrated information from the $\DACP$ measurement. Measuring both observables is crucial for understanding the underlying mechanism of \CP violation.

Besides, localized \CP violation is investigated through measurements of raw \CP asymmetries and triple-product asymmetries across various phase-space regions, with no significant variations observed.

\FloatBarrier
\section*{Conflict of interest}
\noindent The authors declare that they have no conflict of interest.
\section*{Acknowledgments}
%
%
\noindent We express our gratitude to our colleagues in the CERN
accelerator departments for the excellent performance of the LHC. We
thank the technical and administrative staff at the LHCb
institutes.
We acknowledge support from CERN and from the national agencies:
ARC (Australia);
CAPES, CNPq, FAPERJ and FINEP (Brazil); 
MOST and NSFC (China); 
CNRS/IN2P3 (France); 
BMFTR, DFG and MPG (Germany);
INFN (Italy); 
NWO (Netherlands); 
MNiSW and NCN (Poland); 
MCID/IFA (Romania); 
MICIU and AEI (Spain);
SNSF and SER (Switzerland); 
NASU (Ukraine); 
STFC (United Kingdom); 
DOE NP and NSF (USA).
We acknowledge the computing resources that are provided by ARDC (Australia), 
CBPF (Brazil),
CERN, 
IHEP and LZU (China),
IN2P3 (France), 
KIT and DESY (Germany), 
INFN (Italy), 
SURF (Netherlands),
Polish WLCG (Poland),
IFIN-HH (Romania), 
PIC (Spain), CSCS (Switzerland), 
and GridPP (United Kingdom).
We are indebted to the communities behind the multiple open-source
software packages on which we depend.
Individual groups or members have received support from
Key Research Program of Frontier Sciences of CAS, CAS PIFI, CAS CCEPP, 
Fundamental Research Funds for the Central Universities,  and Sci.\ \& Tech.\ Program of Guangzhou (China);
Minciencias (Colombia);
EPLANET, Marie Sk\l{}odowska-Curie Actions, ERC and NextGenerationEU (European Union);
A*MIDEX, ANR, IPhU and Labex P2IO, and R\'{e}gion Auvergne-Rh\^{o}ne-Alpes (France);
Alexander-von-Humboldt Foundation (Germany);
ICSC (Italy); 
Severo Ochoa and Mar\'ia de Maeztu Units of Excellence, GVA, XuntaGal, GENCAT, InTalent-Inditex and Prog.~Atracci\'on Talento CM (Spain);
SRC (Sweden);
the Leverhulme Trust, the Royal Society and UKRI (United Kingdom).

\section*{Author contributions}
\noindent The authors listed in the paper contribute equally to the physics results, during different processes of this experimental measurement. 




\addcontentsline{toc}{section}{References}
\bibliographystyle{LHCb}
\bibliography{main,standard,LHCb-PAPER,LHCb-CONF,LHCb-DP,LHCb-TDR}

\newpage
\centerline
{\large\bf LHCb collaboration}
\begin
{flushleft}
\small
R.~Aaij$^{38}$\lhcborcid{0000-0003-0533-1952},
A.S.W.~Abdelmotteleb$^{57}$\lhcborcid{0000-0001-7905-0542},
C.~Abellan~Beteta$^{51}$\lhcborcid{0009-0009-0869-6798},
F.~Abudin{\'e}n$^{57}$\lhcborcid{0000-0002-6737-3528},
T.~Ackernley$^{61}$\lhcborcid{0000-0002-5951-3498},
A. A. ~Adefisoye$^{69}$\lhcborcid{0000-0003-2448-1550},
B.~Adeva$^{47}$\lhcborcid{0000-0001-9756-3712},
M.~Adinolfi$^{55}$\lhcborcid{0000-0002-1326-1264},
P.~Adlarson$^{85}$\lhcborcid{0000-0001-6280-3851},
C.~Agapopoulou$^{14}$\lhcborcid{0000-0002-2368-0147}
% ...

{\footnotesize \it

% ...
$^{14}$Universit{\'e} Paris-Saclay, CNRS/IN2P3, IJCLab, Orsay, France\\
% ...
$^{38}$Nikhef National Institute for Subatomic Physics, Amsterdam, Netherlands\\
% ...
$^{47}$Instituto Galego de F{\'\i}sica de Altas Enerx{\'\i}as (IGFAE), Universidade de Santiago de Compostela, Santiago de Compostela, Spain\\
% ...
$^{51}$Physik-Institut, Universit{\"a}t Z{\"u}rich, Z{\"u}rich, Switzerland\\
% ...
$^{55}$H.H. Wills Physics Laboratory, University of Bristol, Bristol, United Kingdom\\
% ...
$^{57}$Department of Physics, University of Warwick, Coventry, United Kingdom\\
% ...
$^{60}$School of Physics and Astronomy, University of Glasgow, Glasgow, United Kingdom\\
$^{61}$Oliver Lodge Laboratory, University of Liverpool, Liverpool, United Kingdom\\
% ...
$^{69}$Syracuse University, Syracuse, NY, United States\\
% ...
$^{85}$Department of Physics and Astronomy, Uppsala University, Uppsala, Sweden, associated to $^{60}$\\
% ...
}
\end{flushleft}

\end{document}